# An Improved and Parallel Version of a Scalable Algorithm for Analyzing Time Series Data

Andreas Vitalis*

University of Zurich

Department of Biochemistry

Winterthurerstrasse 190, CH-8057 Zurich, Switzerland

* To whom correspondence should be addressed: a.vitalis@bioc.uzh.ch

**ABSTRACT**

Today, very large amounts of data are produced and stored in all branches of society including science. Mining these data meaningfully has become a considerable challenge and is of the broadest possible interest. The size, both in numbers of observations and dimensionality thereof, requires data mining algorithms to possess time complexities with both variables that are linear or nearly linear. One such algorithm, see Comput. Phys. Commun. **184**, 2446-2453 (2013), arranges observations into a sequence called the progress index. The progress index steps through distinct regions of high sampling density sequentially. By means of suitable annotations, it allows a compact representation of the behavior of complex systems, which is encoded in the original data set. The only essential parameter is a notion of distance between observations. Here, we present the shared memory parallelization of the key step in constructing the progress index, which is the calculation of an approximation of the minimum spanning tree of the complete graph of observations. We demonstrate that excellent parallel efficiencies are obtained for up to 72 logical (CPU) cores. In addition, we introduce three conceptual advances to the algorithm that improve its controllability and the interpretability of the progress index itself.

## 1. INTRODUCTION

Data and their analysis play an increasingly central role in our daily lives. For example, IT companies mine browsing histories to personalize user experiences on their respective platforms. Unfortunately, many data sets are noisy, of high complexity (or dimensionality, denoted $D$), and extremely large. As such, there is a pressing need for scalable algorithms that can handle these properties of the input data [12, 23, 28]. Partially automated visual surveillance provides a useful illustration [3]. In particular, the "live" mining of collected video footage, which are time series of images each with *ca*. $10^5$-$10^7$ pixels, requires algorithms with reasonable cost in $D$ and linear cost in the number of images, $N$. Poor scalability in $D$ may be compensated by sacrifices in resolution, but superlinear scaling with $N$ will quickly lead to a progressive lag of the analysis. These constraints restrict the available algorithmic space tremendously. It can be argued that the sciences face a similar race condition with data generation and storage becoming increasingly automated. In particular, parallelism is more commonly used to boost data generation rather than to mine the results [10]. It is therefore timely to present parallel versions of scalable algorithms in data science as done here.

Many data sets being recorded are time series [11], *e.g.*, video footage, neuronal activity [13], or computer simulations of galaxy formation [24]. Systems on the molecular scale usually behave like memory-free, stochastic dynamical systems on relevant time scales. Computer simulations have contributed greatly to the appreciation of the complexity of these molecular systems, *e.g.*, of proteins in the life sciences [16]. The desire to mine the vast amounts of time series data generated today has inspired the development of a scalable algorithm [4] as follows. It takes as input $N$ observations or snapshots, each of dimensionality $D$, and an expression for computing a continuous measure of pairwise similarity between snapshots. It produces a sequence of these snapshots termed progress index. The progress index can be annotated by features of the data ("structural" annotation) and by dynamical information. The goal is to provide a comprehensive and compact overview of the states populated by the system including their geometrical and dynamical delineations [5]. The strengths of the method are the avoidance of overlap of states, the maximal resolution, which is due to it operating directly at the snapshot level, and its efficiency.

Structural annotations are easy to construct but not easy to select unless $D$ is very small or sufficient domain knowledge exists. The dynamical partitioning is revealed by a more general, cut-based annotation function [4, 17] that provides information about numbers of transitions between sets of snapshots. At any given point along the progress index, these sets are those snapshots to the left (**S**) and those to the right (**A**). The cut-based annotation function makes sense only if the data carry information about sequences of events, *e.g.*, if they are time series data.

The progress index is constructed from left to right, and as such the task is equivalent to the construction of set **S** for any given point. The algorithm proceeds as follows. From a generally arbitrary starting snapshot, which initializes **S** with its first member, $s_i$, one finds the closest available neighbor to any member of **S** that is not already in **S**. The proximity of a neighbor is evaluated based on a measure of similarity, which is the only essential parameter of the method. Thus far, published work has used only geometric similarity measures, most often Euclidean distances in high-dimensional spaces [6, 15]. Once all $N$ snapshots have been added, the sequence can be annotated. At any given point, the cut-based function simply counts the number of direct transitions between the corresponding sets **S** and **A** when taking the data in their original input order. In the language of chemical kinetics, points with low transition counts are generally indicative of free energy barriers, which separate the various free energy basins. If the representation of the underlying dynamical system is appropriate, the latter are characterized by a notably higher sampling density. The cut value at any given point, $c(i)$, is related to the average of the mean first passage times between the corresponding sets **S**(i) and **A**(i) as:

$$\tau_{\mathbf{S}(i)\to\mathbf{A}(i)} + \tau_{\mathbf{A}(i)\to\mathbf{S}(i)} = 2N/c(i) \qquad (1)$$

The annotated progress index produces a compact visualization called a SAPPHIRE plot [5] that reveals a large number of salient features of the complex system.

The method as described thus far corresponds to the exact algorithm in the original paper [4]. There is an approximate version that offers near-linear scalability with data set size. Its approximation is in the way the nearest available neighbor to the set **S** is obtained, and this is described in detail below. In this contribution, we present a shared memory parallelization of the rate-limiting core step of this approximate algorithm and demonstrate it to work perfectly on a modern CPU architecture if the computational cost is limited by FLOPs from pairwise distance evaluations. In addition, we present three conceptual advances as follows. First, we report an improvement of the neighbor search procedure such that it enables the approximate algorithm to approach the solution of the exact algorithm asymptotically using two controllable parameters. Second, we extend an underlying clustering algorithm [26] such that it can provide a better preorganization of the data. Third, we introduce a modification to the creation of the progress index that applies to both exact and approximate algorithms. This modification provides an improved grouping of points belonging to regions of low sampling density. The relation of the improvements to the algorithm is summarized in Fig. 1.

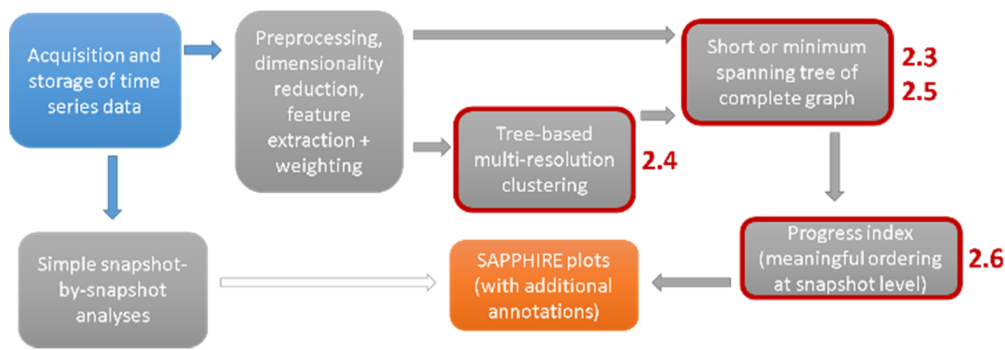

**Figure 1:** Schematic illustration of the data mining pipeline used for obtaining SAPPHIRE plots. Filled arrows indicate essential input. The boxes with bold outlines are those steps for which improvements are presented here. The corresponding section numbers for the manuscript are shown next to the boxes. The two alternative ways of arriving at the spanning tree are the exact (long arrow) and approximate algorithms, respectively.

When possible, we test these modifications for the desired outcome without focusing on the data sets used to perform the tests. In all other cases, we choose examples that require little background knowledge. Because this is a methodological contribution, we present a joint section Methods and Results next, which contains a description of data sets (2.1), a review of the approximate progress index (2.2), the improvements to the neighbor search and clustering (2.3 and 2.4), the parallelization (2.5), and the revision of the progress index itself (2.6). We conclude with a summary and directions for future research (3).

## 2. METHODS AND RESULTS

Version 3 of the open source software package CAMPARI (http://campari.sourceforge.net) hosts all the methods described below and was used to run the tests described.

### 2.1 Data sets

**DS1**: This is a high-dimensional data set from particle simulations that we use to demonstrate improvements to the search procedure. We extract the absolute positions of 83 particles in Cartesian space ($D = 249$), and pairwise distance evaluations for the latter require a 3D alignment procedure [26]. The data set is otherwise small ($N = 10^4$) since we need to compute the (exact) MST.

**DS2**: This is an example trajectory on alanine dipeptide, a very common model system with two essential degrees of freedom, $\phi$ and $\psi$ ($D = 2$). These degrees of freedom are dihedral angles, *i.e.*, they are periodic (circular) variables requiring appropriate corrections [26]. The data are from a molecular dynamics simulation, *i.e.*, kinetic information is contained. The data set is small to facilitate its use as an illustrative example ($N = 10^4$). Due to the low dimensionality, sampling density can be visualized exactly.

**DS3**: This data set, which is taken from published particle simulation work [1], is used for scaling tests, and $N = 1 \cdot 10^7$. Contiguous subsets are used to investigate the behavior of smaller data sets. We extract different representations corresponding either to interparticle distances ($D = 15$) or to the absolute positions of 10 particles in Cartesian space ($D = 30$). Pairwise distance evaluations for the latter require a 3D alignment procedure, which means that they are more costly by a factor of ~50 than the simple Euclidean norm used for the former.

### 2.2 The approximate algorithm

The construction of the sequence defining the progress index can be divided into two parts [4]. Rather than considering all possible neighbors to the growing set **S**(i) at every stage of the algorithm, it is sufficient to consider only those neighbors that would also be neighbors in the appropriate minimum spanning tree (MST). This is the MST of the complete, undirected graph of all snapshots, $G_S$, which has $N(N-1)/2$ edges and $N$ vertices. The edge weights of $G_S$ are defined as the pairwise distances between snapshots. Thus, the progress index can be obtained by first calculating the MST, which is computationally expensive, followed by a cheap operation of deriving the actual sequence from the spanning tree. Several dedicated algorithms exist to find the MST of a graph. In Borůvka's algorithm [21], it is obtained through successive merging operations of subtrees (acyclic subgraphs). Starting with $N$ subtrees all of cardinality 1, consecutive merging operations reduce the number of subtrees in exponential fashion. In each such operation, a single candidate edge per vertex is identified. Candidates are the shortest available edges between subtrees that involve the vertex in question. Duplicated edges and edges creating cycles are pruned, and the remainder become part of the MST. The procedure is repeated until there are exactly $N-1$ edges.

This algorithm is amenable to the following simplification, which is critical for scalability. Instead of identifying exactly the shortest eligible edge per vertex, we make a limited number of random guesses from a preprocessed pool of candidates. The resultant spanning tree will no longer be minimal and is denoted as a short spanning tree (SST). The number of guesses, $N_g$, is an important parameter and is expected to remain constant with $N$ [4]. This is only possible if the preprocessing is able to overcome the decreasing chance of randomly finding nearby points as $N$ grows. We accomplish this by a preliminary clustering using a tree-based algorithm [26] introduced recently (to avoid confusion, we emphasize that this is a different tree than the ones mentioned above for Borůvka's algorithm). The clustering algorithm, which operates with near-linear time complexity itself, partitions all snapshots into clusters for a total of $H+1$ levels of increasing resolution. At the root (the $0^{th}$ level), all snapshots are in a single cluster. The $H$ actual levels use distance thresholds ranging from $d_1$ to $d_H$. The leaf level offers the finest resolution and provides most of the information that is exploited for the construction of the SST. Borůvka's algorithm in conjunction with a fixed number of candidate edges per vertex gives the desired time complexity of $O(N \log N)$.

The approximate construction of the progress index implemented the limited neighbor search as follows: for a vertex $s_i$, a maximum of $N_g$ guesses of near neighbors are drawn from a pool of eligible vertices provided by the cluster $c_k^h$. Here, $c_k^h$ is the cluster at hierarchy level $h$ that contains $s_i$. The value of $h$ is determined as the highest value for which the respective cluster fulfilling $s_i \in c_k^h$ offers at least one eligible candidate. This procedure places a strong emphasis on the structure of the clustering tree. For example, it is initially impossible for two snapshots in different leaf-level clusters, which are not singles, to be joined as part of the SST. This led to the restriction that the SST could not generally approximate the MST to arbitrary precision (see Fig. S.5 in [4]). The first improvement we present here exposes this limitation as unnecessary.

### 2.3 Better search exhaustiveness and controllability: Improvement of the SST

Here, we propose to make the number of guesses, $N_g$, be enforced rigorously whenever possible. For vertex $s_i$, let $c_k^h$ be the finest (highest in the tree) cluster containing $s_i$ that offers at least one eligible candidate. If the number of candidates exceeds $N_g$, no adjustments to the algorithm are made. If $N_g$ is larger, however, the tree is descended and cluster $c_k^{h-1}$ is used to offer a larger pool of candidates. This procedure is continued until either the number of candidates considered exceeds $N_g$ or until a user-defined threshold, $\sigma_{max}$, for the number of additional tree levels to use has been exceeded. A caveat is hidden in the fact that the vertices scanned in $c_k^{h-1}$ can partially be the same as in $c_k^h$ and above. We currently do not address this redundancy.

The introduction of the parameter $\sigma_{max}$ and the resultant increase of the search space ensure that the SST can approximate the MST to an arbitrary degree. Fig. 2 proves this point. In Fig. 2(A), the identity in terms of shared edges between MST and the SSTs is shown to increase dramatically even for small values of $\sigma_{max}$. The original algorithm essentially corresponds to the case with $\sigma_{max} = 0$. The available benefit depends on $N_g$, which is expected. The preorganization continues to strongly determine the result as indicated by the tiny differences between different SSTs obtained with the same settings. Fig. 2(B) emphasizes that the net length of the SST is within 5% of the length of the MST for all values. For application in the progress index, this is a more representative measure of quality. As a boost to efficiency, we implement an additional feature that stores and maintains a fixed-size list of 5 prior guesses per vertex for possible reuse in later stages of Borůvka's algorithm [21]. This widens the candidate pool whenever the search is random for all stages of the algorithm but the first.

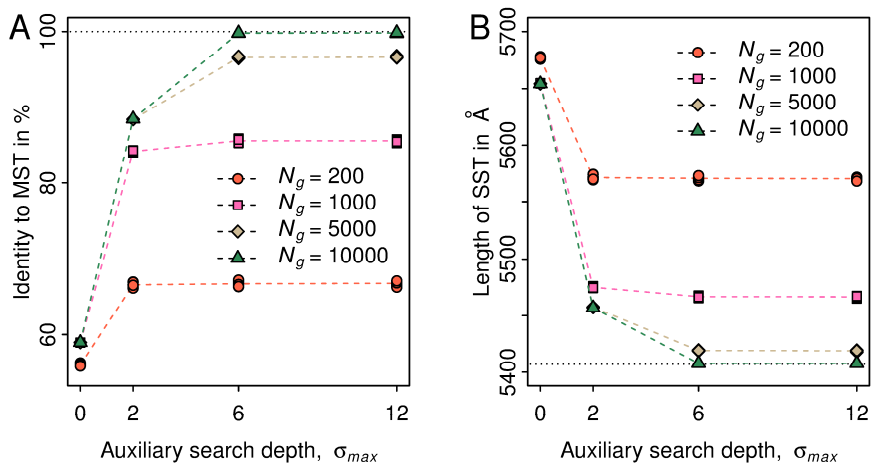

Figure 2: Comparison of SSTs and the MST for a subset of DS1 as a function of $N_g$ and $\sigma_{max}$. The parameters $H$, $d_1$, $d_H$, and $\eta_{max}$ for the tree-based clustering algorithm were 12, 3.5Å, 0.75Å, and 10, respectively. For every set of values of $N_g$ and $\sigma_{max}$, at least 6 independent trials were performed. The variability is generally so low that symbols overlap. A. The identity to the MST, which is computed as the percentage of shared edges, is plotted for the various SSTs. Lines are drawn purely to guide the eye. B. The same as A for the net length of the SSTs.

For the parameter $\sigma_{max}$ to be most effective, the tree structure should be such that there is a more or less constant factor in the numbers of snapshots per cluster between adjacent tree levels. This can be tuned by the thresholds $d_1$ to $d_H$ but a simple linear interpolation has sufficed in most applications until now. To ensure that the candidate pools are as useful as possible, the groupings at all levels should be free of outliers and cluster overlap, and this is discussed next as the second improvement.

### 2.4 Better preorganization: Improvement of the tree-based clustering

In the published version [26], the tree-based clustering calculates a grouping primarily at the leaf level while the levels closer to the root provide scalability. Scalability arises because the hierarchical tree offers parent-child relations that provide a fixed search space (not to be confused with the search space considerations above) used to group snapshots into clusters. The algorithm uses a two-pass approach. In the first pass, the tree is constructed for all levels except $H$. In the second pass, the (now fixed) tree is used to calculate the grouping at the leaf level ($H$). The groupings at levels below $H$ are of inferior quality because the tree structure was still evolving as snapshots were added.

Here, we straightforwardly extend this two-pass idea to a multi-pass approach. Descending from $H$ downward, the corresponding level is first deleted. This is followed by a scan across all snapshots using all lower levels to create the grouping. For level $H$ - 1, this means levels 0 to $H$ - 2. Higher levels are ignored during this procedure. Thus, level $H$ - 1 is created in exact analogy to the way level $H$ was created. The procedure continues for $\eta_{max}$ levels, which is a controllable parameter with a maximum value of $H$ - 2. At very coarse levels, the additional pass through the data has little to no impact. In addition, we have developed a shared memory parallelization of this algorithm to be presented elsewhere.

Fig. 3 shows a visualization of clusters for a two-dimensional, illustrative data set (DS2). We chose a tree with $H = 8$ and size thresholds spaced evenly between 2.5° ($d_8$) and 100° ($d_1$). We show a comparison at two different levels with thresholds of 16.43°, Figs. 3(A)-(B), and 58.21°, Figs. 3(C)-(D). The multi-pass approach (right-hand side) clearly improves the homogeneity of clusters in terms of size and overlap when compared to the tree-based clustering generated directly in the first pass (left-hand side). In general, the additional passes through the data tend to reduce the numbers of clusters at intermediate levels. The improvement to the hierarchical grouping can be important to any application of the algorithm. We mention the efficient construction of multi-resolution networks as an example. In the context of the progress index, the modification's primary virtue is to make clusters more consistent and thereby predictable for the user.

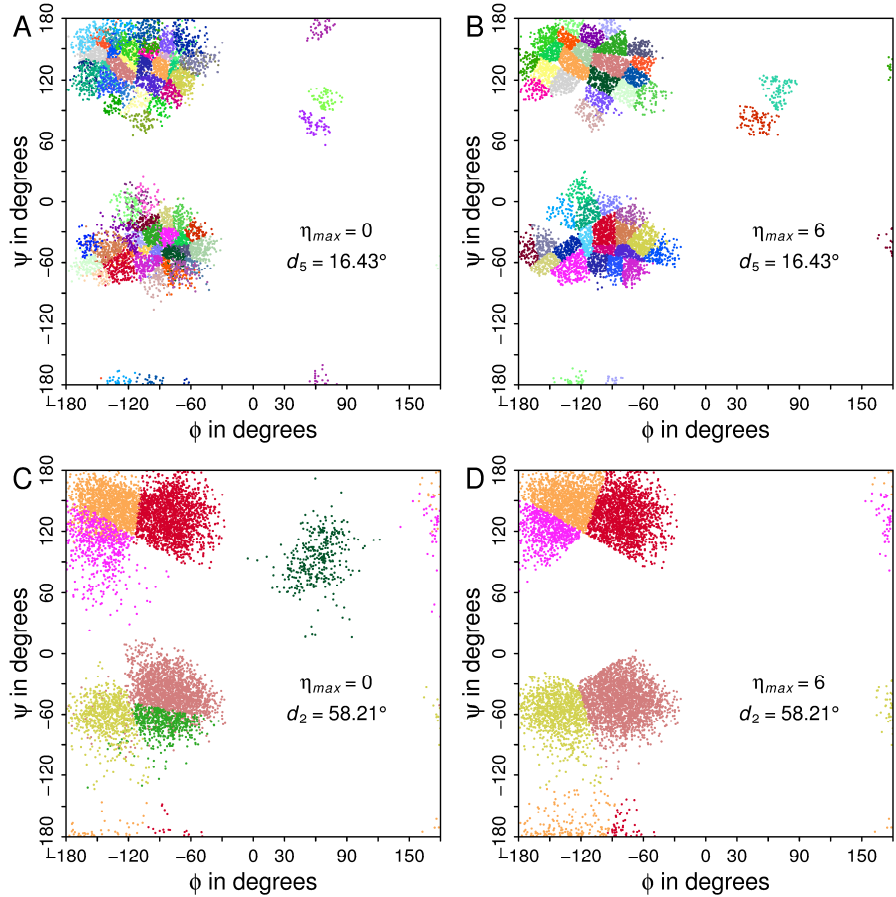

**Figure 3:** Clustering of DS2 at intermediate resolution levels. Settings are given in the text. A. For a cluster size threshold of 16.43°, we show the data in the original space, which are periodic variables, $\phi$ and $\psi$. Dots are colored by cluster membership. The largest clusters encompassing at least 75% of the data are shown (64 clusters). B. The same as A after applying the multi-pass approach (44 clusters). C. The same as A for a size threshold of 58.21° (7 clusters). D. The same as C after applying the multi-pass approach (5 clusters). Colors are all unique but some are bound to be similar. Resolution limitations mean that dots can overlap.

### 2.5 Parallelization of SST construction

We describe here the implementation and performance of a shared memory parallelization of the construction of the SST. The parallelization of the tree-based clustering algorithm will be presented elsewhere. Other elements of the progress index generation such as the computation of the actual sequence from the SST or of the cut functions are not currently parallelized.

The parallelization takes advantage of the fact that every stage of Borůvka's algorithm [21] requires a fixed number of guesses of near neighbors to be performed for every vertex. This means that the total load per stage is roughly $N \cdot N_g$, which we divide at the level of vertices. Every distance evaluation between pairs of snapshots has the same cost in FLOPs. The fundamental assumption is that this cost is performance-limiting. In this case, we expect the parallelization described below to be perfectly balanced.

Schematically, the parallel algorithm for *T* threads for a single Borůvka iteration looks as follows. We use the term "subtree" for the tree structures being merged in Borůvka's algorithm and the term "tree" for the hierarchical tree obtained from the clustering:

For the assigned chunk of clusters across all tree levels $\{c_k^x\}$ do:

  Sort the list of vertices in $c_k^x$ by subtree membership. **(1)**

For assigned chunk of *N/T* vertices $\{s_i\}$ do:

  Initialize the number of guesses for given vertex ($g_i$) to 0 and the tree level ($h_i$) to *H*. **(2)**

  While $g_i$ is less than $N_g$, do:

    Determine the number of eligible candidates in $c_k^{h_i}$. **(3)**

    If the number of candidates is larger than $N_g - g_i$, do:

      For $N_g - g_i$ attempts do:

        Pick a set of random elements in $c_k^{h_i}$ that are not in the same subtree as $s_i$. **(4a)**

        Compute the distance to $s_i$ and, if needed, update the list of nearest neighbors per vertex. **(4b)**

        Increment $g_i$. **(4c)**

    Else do:

      For every element in $c_k^{h_i}$ that is not in the same subtree as $s_i$ do:

        Compute the distance to $s_i$ and, if needed, update the list of nearest neighbors per vertex. **(5a)**

        Increment $g_i$. **(5b)**

    Decrease $h_i$. **(6)**

    Exit if $h_i$ is -1 (all levels including the root have been exhausted). **(7)**

For assigned chunk of *N/T* vertices $\{s_i\}$ do:

  Find the shortest available edge per subtree and write it to a temporary list. **(10)**

For every subtree do:

  Update shared candidate with better candidate from temporary list (requires exclusive handling by a single thread at a time) if available. **(11)**

If thread is master thread then:

  Merge subtrees based on the remaining candidate edges using child-parent pointers. **(12)**

Else:

  Wait. **(13)**

Repopulate temporary and thread-local subtree objects. **(14)**

For assigned chunk of *N/T* vertices $\{s_i\}$ do:

  Update subtree membership. **(15)**

  Manage list of near neighbors by eliminating those that have become ineligible. **(16)**

Synchronize threads. **(17)**

**Scheme 1: Pseudocode for a single thread during a single Borůvka iteration. Indentation is for conditionals and loops, and numbers in parentheses are used to reference specific statements.**

For a scalable implementation even the simplest steps in Scheme 1 must not exceed a constant number of operations per vertex. Step (3) is tricky in this regard, and it relies on the sorting performed in step (1). Rather than having to repeatedly scan a list of uncontrollable size (in O(*C*) time where *C* is the size of the cluster), the sorting by tree membership upfront allows rapid identification of the set of eligible candidates and its size given that the range of entries with the same tree membership can be found by binary search in O(log *C*) time.

The parallelization assumes that the cost of branches (4) and (5) is equivalent. This need not be the case in practical situations for three main reasons. First, the cost of random number generation can be significant relative to the cost of evaluating pairwise distances. Second, compiler-enabled runtime optimizations may differ between (4) and (5). Third, memory and cache management and access costs are likely to be higher if data access is generally noncontiguous as in (4). While we ignore the second reason, the first and third are addressed as follows. We limit the use of random numbers in step (4a) by defining a rigid schedule with respect to the randomly chosen reference point in the set of eligible candidates. This set is derived from the snapshots in a cluster. In all but the first Borůvka stage, the schedule is a stretch of 150 consecutive eligible cluster members. In the first Borůvka stage, it is a set of 150 evenly spaced points that span the entire cluster. Since the reference point is chosen with uniform probability, the resultant picking probability for all snapshots remains uniform as well. With this technique, the demand for random numbers can be made so low that it remains feasible to use a shared random number generator, which requires thread synchronization. The use of consecutive cluster members helps with cache and memory performance even though the list of snapshots in a cluster does not necessarily map to a contiguous stretch in the main data array. Diagnosing this behavior is unlikely to be insightful as it depends very strongly on architecture, compiler, optimization settings, thread affinity model, system load, and the specifics of the data set.

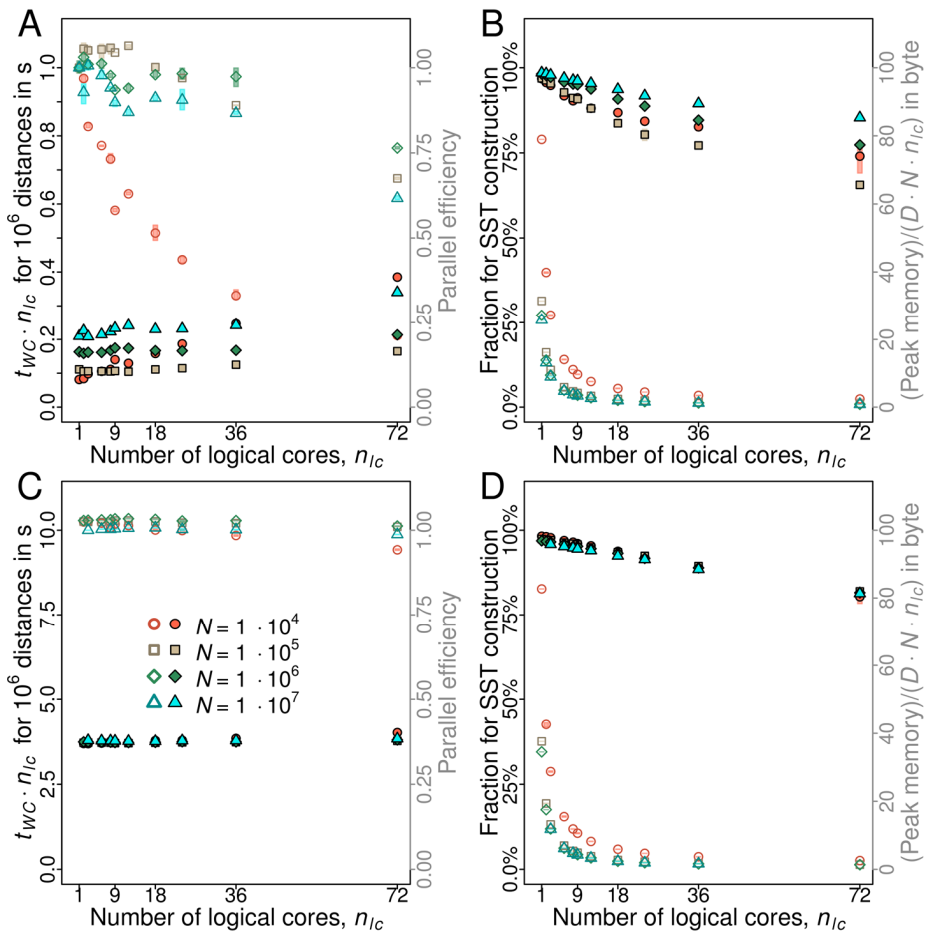

**Figure 4: Performance of the shared-memory parallel algorithm using OpenMP. Code was compiled using the Cray Fortran compiler, version 8.5.5, and recommended optimization settings for the Broadwell architecture. Input data are taken from DS3. The parameters $H$, $d_1$, $d_H$, $\eta_{max}$, $N_g$, and $\sigma_{max}$ were 8, 6Å, 1.5Å, 6, 500, and 7, respectively. Since the number of required Borůvka stages can vary stochastically, all wall-clock times ($t_{WC}$) were normalized by the total**

**number of distance evaluations. Min/max ranges from 2-4 measurements are indicated as semi-opaque rectangles of variable height: they are almost always so small that they vanish behind the plotting symbols. A. Normalized performance and parallel efficiency for a simple distance function with $D = 15$. Open symbols give parallel efficiency (right y-axis) using a single thread as reference, and filled symbols denote normalized times (left y-axis). The measured times include the entire SST construction but not the tree-based clustering or the progress index generation. B. Fractional time cost (filled symbols) and peak memory use (open symbols) for SST construction. The total time considered includes the complete workflow except file I/O. Peak memory has not been optimized and is measured at the OS/process level. C. The same as A for an expensive distance function with $D = 30$. For the largest data set, values for 1 and 2 cores are missing due to runtime limits (for this case only, the performance reference is the mean result for 3 threads). D. The same as B for the data used in C.**

Fig. 4 demonstrates two application examples. The testing architecture is a single compute node on the supercomputer Piz Daint at CSCS, which is a dual-socket machine with two 18-core Intel Broadwell CPUs (Xeon® E5-2695 v4) with hyperthreading enabled (up to 72 threads, which is the same as the physical number of cores on latest generation Knights landing multicore CPUs). By fully loading machines, pinning, for each process, $T$ threads to $n_{lc}$ individual and unique logical cores, and restricting processes to a single socket whenever possible, we obtained representative benchmarks. Fig. 4(A) shows a challenging case. Here, the Euclidean distance is used for a data set with $D = 15$ derived from DS3. The overall cost is very low, *i.e.*, the SST for $10^7$ snapshots is obtained in 4-5 minutes with 72 threads. The simple distance function means that vectorization can be exploited fully in the innermost kernel. As a result, the calculation is not limited by FLOPs but instead by main memory and cache operations. Despite this, parallel efficiency remains above 80% for all but the smallest data set, which is processed in milliseconds in almost all cases. Cache limitations become apparent by the distinct drop in efficiency for the only calculations spanning both sockets, *viz.*, with 72 threads. The main bottleneck appears to be the decreasing likelihood that portions of the main data array loaded into cache can be reused in consecutive iterations of step (4). Cache efficiency and the demand for random numbers depend on the data and can differ between threads, and this gives rise to load imbalance. Fig, 4(B) demonstrates that SST construction remains the performance bottleneck in all cases. It also shows that the peak memory usage fulfills the desired and expected criteria of a) scaling linearly with data set size from $N = 10^5$ onwards; b) not introducing detrimental overhead form the shared memory parallelization (the scaling with the number of threads is also linear but starting from a large value for $n_{lc} = 1$).

Figs. 4(C)-(D) give results for a different feature extraction from DS3 with $D = 30$, which does, however, use an expensive operation in the distance computation (3D alignment, giving rise to a strong increase in cost per distance relative to Fig. 4(A)). Here, performance is limited by FLOPs, and load balance is achieved under all settings. Performance results are extremely consistent across data set sizes spanning 3 orders of magnitude and using 1-72 threads because memory and cache access do not become limiting. These results are also insensitive to most implementation details for the critical step (4a). This includes the choices to draw either as many or much fewer random numbers as there are guesses, to use a shared or a thread-private random number generator, and to enforce the scanning of randomly selected stretches of consecutive cluster members.

We conclude the performance description by emphasizing that feature extraction and distance function are completely modular entities with respect to the parallelization. In particular, it is simple to write architecture-specific kernels for both, which can take maximal advantage of available vector extensions. This two-level parallelization mimics the design of the hardware itself. As shown, with a cheap distance function as in Fig. 4(A), the calculation is unfortunately not limited by FLOPs. This makes it much harder to predict performance when changing $D$ and $N$ [14] and reduces possible gains from hardware-specific vectorization.

## 2.6 An improved progress index

With the spanning tree in hand, the progress index construction itself is a cheap operation. This is irrespective of whether the MST or a SST are used. From an arbitrary starting point, it will add vertices based on the shortest available edge to any vertex already added. In the worst case, it requires the sorting of a list of *N* - 1 edges by weight (the pairwise distances between snapshots). Due to the rigorous sorting by distance, the sequence tracks local sampling density extremely well. However, the sorting by distance also implies that all regions of low sampling density are treated the same way. Whenever the shortest available distance becomes as large as the mean edge length in transition regions, all the points in adjacent low density sampling regions not having been added yet become eligible. This includes what can be called "fringe" regions of basins, *i.e.*, outliers that do not connect to anything else. Many of these outliers will be connected by an edge in the spanning tree that is longer then the edge required to hit upon a new region of high sampling density. This causes the outliers to be added much later, often at the very end. This aspect of the protocol creates stretches in the progress index that lack any notion of locality.

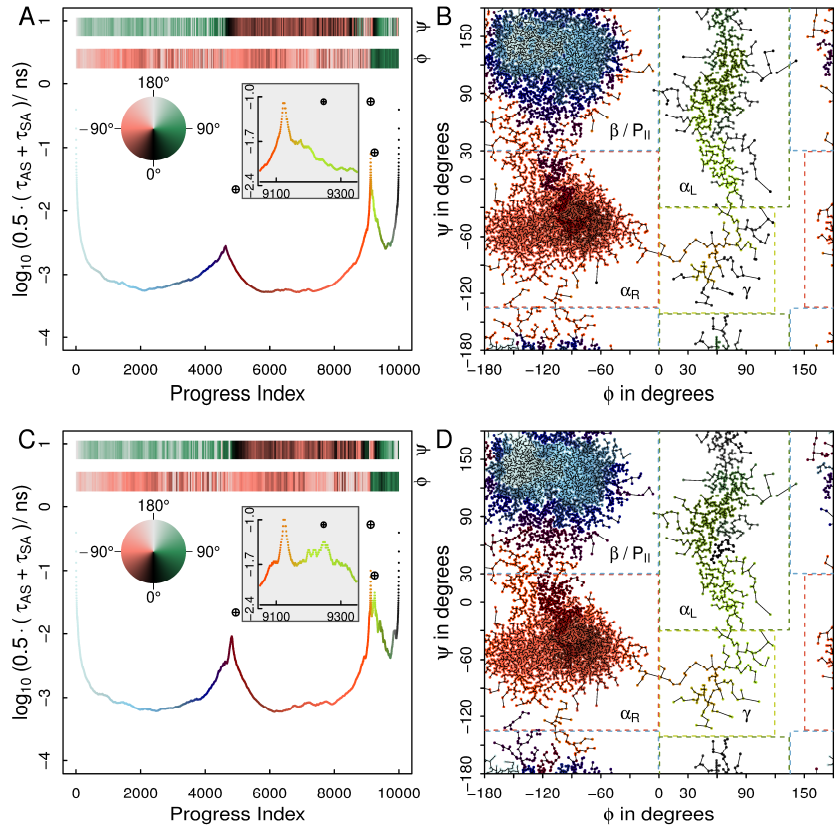

**Figure 5: Illustration of the improved progress index construction. A.** The progress index for DS2 is plotted using the standard cut-based annotation function and two structural annotations corresponding to the two degrees of freedom in the representation. These data are based on the MST and use a starting snapshot in the β-basin. The color wheel provides the legend for the structural annotations. Points derived from a 4-state Markov model are highlighted with the "crosshair" symbols. Their x-coordinates correspond to the cumulative probabilities of the β/$P_{II}$, $α_R$, and γ basins (in this order). The characteristic times are taken as the inverse of the total number of transitions into the state immediately to the right from any state to the left. This is consistent with the 2-state model underlying the cut function. The parameter $ρ_f$ is set to 0. The inset highlights the region where the γ basin should be found. **B.** Individual data points are plotted in the original data space and colored by their position in the progress index (see A for color code). Thin black lines are used to plot the MST. The labeled rectangles indicate regions used to construct the 4-state Markov model mentioned above and in the text. Both variables are of course periodic which affects lines and rectangles. **C.** The same as A for $ρ_f$ = 16. **D.** The same as B for $ρ_f$ = 16.

Figs. 5(A)-(B) illustrate this problem for the simple 2D data set (DS2). Fig. 5(A) shows the progress index with the cut-based annotation function, which reveals 3 major basins. The β/$P_{II}$, $α_R$, and left-handed basins are visited in this order when starting from a snapshot in the β region (see structural annotation). In Fig. 5(B), the points of the trajectory are colored by their position in the progress index, which confirms the result in Fig. 5(A). In addition, the MST of the data is shown. It is clearly seen that the sampling density in the transition region between β/P$_{II}$ and α$_R$ is at least as high as that in the fringe regions of both basins. This leads to the undesired effect that points in these fringe regions fail to be added until positions of the progress index of 8500 and above. DS2 is of low enough dimensionality that we can define basins also by visual inspection (see rectangles in Fig. 5(B)). By coarse-graining the original trajectory geometrically, we can derive populations and approximate transition rates from a 4-state Markov model. If the progress index is perfect, the cut-function should pass through the resultant 3 points highlighted in Fig. 5(A). Instead, all transition rates are clearly overestimated, and the delineation of the α$_L$ and γ basins is missed even qualitatively (see inset).

The final modification proposed here tries to address these shortcomings at the level of the spanning tree. Given that outliers in fringe regions do not connect to anything else, it is reasonable to assume that the progress index becomes more intuitive if they are included at the same time as the parent basin itself. Unlike transition regions between basins, the fringe regions of basins are very likely to correspond to terminal (leaf) vertices in the SST. The information as to whether a vertex fulfils this criterion is readily available from the SST. We propose to simply "fold back" such terminal vertices onto their parent. In specific terms, this means that all neighboring vertices, which are available to the current set **A** in the SST, are sorted by increasing distance in two separate subsets: those that are leaf vertices and those that are not. The subset of leaf vertices is categorically processed first. This has the consequence that outliers appear in close proximity to the snapshot they are nearest to in the progress index. More and more snapshots can be added to the leaf subset by applying the procedure repeatedly. For this, the current members of the leaf subset are ignored when scanning the SST for additional leafs. We term the number of these cycles $\rho_f$. In terms of the original SST, after $\rho_f$ operations, the neighbors in the leaf subset of a given point are all in terminal branches of maximum length $\rho_f$.

The result of this procedure is shown in Figs. 5(C)-(D). Fig. 5(D) clearly shows that the points in the fringe region of the β and α$_R$ basins now appear in the progress index within their respective basins. This improves the delineation of the two major basins, and the cut function in Fig. 5(C) gives a barrier position and height that is much closer to the expectation derived from the 4-state Markov model. Even though it is only constituted by ~100 snapshots, the revised progress index also identifies the γ basin as kinetically distinct (see inset in Fig. 5(C)). Importantly, all of the conclusions that could be gained from Fig. 5(A) are also available from Fig. 5(C). This clearly suggests the use of nonzero values for $\rho_f$ as a systematic and reliable improvement.

The example system used in Fig. 5 is of very low dimensionality. In high-dimensional spaces, the structure of the spanning tree can change. Specifically, it is likely that the fraction of leaf nodes increases, and that the maximum path length decreases. This means that small values of $\rho_f$ can already have a significant impact.

## 3. DISCUSSION AND CONCLUSIONS

In this contribution, we have shown improvements to a recently developed analysis scheme for complex and large data sets with a focus on time series data describing stochastic dynamical systems (Fig. 1). The approximate algorithm to construct the progress index, which scales with O($DN\log N$), includes as its computationally expensive core step the construction of the spanning tree of the complete graph of snapshots, $G_S$. We have demonstrated that our strategy of using a multi-resolution clustering to preorganize the data can be improved upon to approach the MST asymptotically by tuning simple parameters (Fig. 2). The multi-resolution clustering is subject to a systematic extension itself (Fig. 3). The performance of the OpenMP shared memory parallelization was shown to be very good even under unfavorable conditions (Fig. 4(A)) and perfect under favorable conditions (Fig. 4(C)). Lastly, we solve a specific problem in the progress index construction that could lead to a loss of information content caused by a poor delineation of regions of low sampling density (Fig. 5).

When dealing with very large and complex data sets, much of the analyses being performed are guided by intuition and prior experience of researchers. In data sets with structurally equivalent samples such as images, clustering [27] is probably the most common unsupervised technique to obtain an initial overview. Our scalable tree-based clustering algorithm, extended here to provide a consistent, multi-resolution picture, is a useful tool in this regard. For dynamical systems, it can also be exploited to construct predictive network models. Of course, clustering requires choosing a measure of similarity between samples, and this is often a difficult decision in and of itself [18, 19]. The progress index is a scalable technique yielding a compact visualization that is not only useful in understanding the complex data set but also in first evaluating the suitability of a chosen measure of similarity, which often implies a feature extraction from the data. The resultant overview can be exceptionally rich as was documented recently [5, 6]. Our parallelization means that data sets with millions of snapshots can be handled on a time scale of minutes to hours on modern multicore CPUs, which is essential if tuning of the similarity measure is required. The main limitation of the current implementation is memory, which scales linearly with both $D$ and $N$ (see Figs. 4B and 4D). We are currently working toward a distributed memory implementation that will add MPI as a third layer of parallelization. While this may entail considerable overhead, we anticipate that scalability for very large data sets can be achieved.

Dimensionality reduction [25] and blind signal separation techniques [8] are commonly used in unsupervised learning to generate features. For example, 2D histograms of automatically generated projections of the input data, *e.g*., from diffusion maps [9] or time-lagged independent component analysis [20], serve as vehicles to visually communicate information about the structure of a complex data set. The relative quality of different projections is hard to classify [2] and depends obviously on the "true" dimensionality of the data. Projections onto predefined collective variables such as those used in enhanced sampling techniques [22] are particularly prone to misrepresentation on account of overlap [17]. Due to its efficiency and resolution, the progress index can indeed be used to investigate the effects of a progressive reduction in dimensionality. This is achievable by obtaining a series of cut-based annotations to reveal how states are lost upon decreasing the number of retained features. The cut annotation of the progress index (see Fig. 5(A)) has the disadvantage that the barriers, even if quantitatively accurate, are barriers between sets of states. However, semiautomatically extracted sets of states can be combined profitably with Markov state models [7] to have access to a data-derived low resolution model of the complex system, which should allow predictions of its long-time dynamics. This joint methodology is an active area of research.

A shared memory parallelization of the clustering algorithm [26] is required to make the bulk of the entire workflow parallel and has been utilized for the data in Fig. 4 (to be presented elsewhere). Ongoing work is also concerned with the automatization of the structural annotations of the data in SAPPHIRE plots. The latter would make sure that an informative visualization is available with minimal user interaction also for data sets that are not time series data.

## 4. ACKNOWLEDGMENTS

The author acknowledges Dr. Amedeo Caflisch for helpful discussions and comments on the manuscript and Dr. Nicolas Blöchliger for performing tests. This work was supported by a grant from the Platform for Advanced Scientific Computing (PASC project "Scalable Advanced Sampling in Molecular Dynamics").